\def\Roman#1{\uppercase\expandafter{\romannumeral#1}}
\documentclass[a4paper,12pt]{article}
\usepackage{cite}
\usepackage{amssymb,amsfonts}
\usepackage[mathscr]{eucal}
\usepackage{mathrsfs}
\usepackage[utf8x]{inputenc}

\title{Dependent coordinates in the Lagrange-Poincar\'{e} equations for mechanical systems with  symmetry}

\author{S. N. Storchak}

\author{S. N. Storchak\footnote{E-mail adress: storchak@ihep.ru}\\
\small{Institute for High Energy Physics, Protvino, Moscow Region,142284,Russia}}

\begin{document}

  \maketitle

\begin{abstract}
The Lagrange--Poincar\'{e} equations for the  mechanical system describing the motion of a scalar particle on a Riemannian manifold with a given  free and  isometric action of  a compact Lie group  is obtained. In an arising principle fibre bundle, the total space 
of which serves as a configuration space of the considered mechanical system,  the local description  of the reduced motion is done 
in terms of  dependent coordinates. In obtaining of   the equations we  use  the variational principle  developed by Poincaré  for 
the mechanical systems with a symmetry.
\end{abstract}

\section{Introduction}
 The main methods used at present for the study of  mechanical systems with symmetries are based on the reduction theory \cite{AbrMarsd,Marsden}.
The theory gives us a necessary  instrument for revealing an existing internal  motion   in  the  original system. 
This is achieved by ``removing''  symmetry out  of the system which leads to a new mechanical system defined on the reduced space.

In the reduction theory, the dynamical behaviour  of the system is described with the help of the Lagrange--Poincar\'{e} equations (the reduced Euler--Lagrange equations) \cite{Marsden_1}. In this system of equation consisting of two equations,
the first   equation, known as the ``horizontal equation'', represents the local evolution given on the reduced space.
The second equation is related to the local evolution of the group variable (or more precisely, to the evolution of the variable in the Lie algebra of the symmetry group.)

The study of the reduction in mechanical systems, in addition to their own interests, are motivated by the possibility to use the obtained methods  in  other dynamic systems, for example, such as those that are associated with  the mechanics of  fluids, with  the various  field-theoretical models and etc.
Of course, it should be done with some care since we are trying to apply the methods obtained for the finite-dimensional systems to the systems with  infinite-dimensional degrees of freedom.

     As for the field theory,
we are very interested in methods borrowed from the finite-dimensional dynamical systems   
which can  be used to study  the reduction in the various models of gauge fields.

It is known that in mechanics there is  a simple finite-dimensional dynamical  system which can  serve for this purpose. 
This  system represents a classical motion of the scalar particle on a smooth (compact) Riemannian manifold on which a free  proper the isometric smooth action of the (compact) Lie group is given. Due to the symmetry, the configuration space of this mechanical system, an original Riemannian manifold, can be viewed as the total space of the principal fibre bundle. This principal bundle carries the natural connection known by the name of the ``mechanical'' connection.

Although the Lagrange--Poincar\'{e} equations for this system have been   obtained earlier in \cite{Marsden_2,Cendra},  but one of the main questions, which is  important for the gauge field    theories,  was not considered in this paper. 
The question is related to the description of the evolution of the system in the reduced space. In gauge theory, the evolution of the gauge fields on the orbit space (on the base of the principal bundle) can not be represented explicitly. This evolution is described  with the help  of the dependent coordinates given on appropriate gauge surface. (This surface is determined  by a chosen  gauge.)

However, in the cited papers the possibility of using  dependent coordinates in the Lagrange--Poincar\'{e} equations was not  considered.
Note that in the equations of Lagrange-Poincaré obtained for field theories in \cite{Ratiu}, this aspect of the reduced evolution also was not investigated.

In this paper, our aim is to clarify the issues arising in a local description of the reduced motion in terms of the dependent coordinates in a mechanical systems with a symmetry and to get the Lagrange--Poincar\'{e} equations for the   mechanical system which is mentioned above. In obtaining of this equations we are based on the variational methods developed by Poincaré  for the mechanical systems with a symmetry.

The plan of the paper is as follows. In  Section 2 we give a short introduction into the geometry of our problem. Section 3 is devoted to the derivation of the Lagrange--Poincar\'{e} equations. In the last Section we discuss the obtained result together with some of it possible applications. In Appendix we derive the relation between the partial derivatives of the velocities and the deformations that are necessary for the variational calculus.

\section{The principle fibre bundle coordinates}
We consider a motion of scalar particle on a smooth (compact) finite-dimensional Riemannian manifold $\mathcal P$. It is assumed that there is  a free, proper, isometric and smooth action  of a compact Lie group $\mathcal G$ on this manifold. 
  This right action can be written in coordinates as ${\tilde Q}^A=F^A(Q{}^B, a^{\alpha})$, where  $Q^A, A=1,\ldots,N_{\mathcal P},$ are the coordinates given on $\mathcal P$ and $a^{\alpha}$, $\alpha= 1 \ldots N_{\mathcal G}$, are the coordinates of a group element. 

From the general theory it follows that the original manifold $\mathcal P$ has structure of the total space of a principal bundle $\pi: \mathcal P\to\mathcal P/\mathcal G=M$. This means that  one can introduce the local coordinates connected  with the coordinates of the fibre bundle on the original manifold  $\mathcal P$. 
To perform this we will follow to the methods given in our papers \cite{Storchak_1,Storchak_11,Storchak_12, Storchak_2}. 
These methods are generalization of those that have been proposed for  finite-dimensional systems in \cite{Razumov} and  gauge field theories in  \cite{Creutz}.

In accordance with these methods the coordinates can be introduced as follows.
The bundle coordinates  are given by the use of the local sections, which in our case are determined by a local  ``gauge'' surface $\Sigma$, a local submanifold of $\mathcal P$. The submanifold $\Sigma$ is given by a set of equations $\chi^{\alpha}(Q)=0, \alpha= 1,\ldots N_{\mathcal G}$. It is required that this local submanifold has a transversal intersection with  the orbits of the group $\mathcal G$ in $\mathcal P$. 
The set of the coordinates $Q^A,A=1,\ldots,N_{\mathcal P}$,  satisfying  the equations $\chi^{\alpha}(Q)=0$ are called  the dependent coordinates.  They are denoted by  $Q^{\ast}{}^A$ (that is, we have  $\chi^{\alpha}(Q^{\ast})=0$).

The dependent coordinates $Q^{\ast}{}^A$ together with the group coordinates $a^{\alpha}$ are used to set the  coordinates of an arbitrary point $p$ given on the principal bundle. If a point $p$ has a coordinates $Q^A$, then the coordinates $a^{\alpha}(Q)$ are determined from the equation
\[
 \chi^{\alpha}(F^A(Q, a^{-1}(Q)))=0.
\]
After that, the coordinates $Q^{\ast}{}^A(Q)$ can be found by moving the point $p$ to the submanifold $\Sigma$ : $Q^{\ast}{}^A=F^A(Q, a^{-1}(Q))$.

Note that since the trivial principal bundle $\Sigma\times \mathcal G \to \Sigma$ is locally isomorphic to the principal fibre bundle $P(\mathcal M,\mathcal G)$, we can also use the dependent coordinates $Q^{\ast}{}^A$ for description of the motion given on the orbit space $\mathcal M$.
 
The interconnection between the coordinates $Q^A$ and $(Q^{\ast}{}^A, a^{\alpha})$ for the same point  gives  us  a rule by which we can perform the replacement of the coordinates on the original manifold $\mathcal P$. 
As a consequence of this replacement we have the following transformation of the coordinate vector fields:
\begin{eqnarray*}
\frac{\partial}{\partial Q^{B}}&=&F^{C}_{B}
(F(Q^{\ast},a),a^{-1})N^{A}_{C}(Q^{\ast})
\frac{\partial}{\partial Q^{\ast}{}^A}\nonumber\\
&&+F^{E}_{B}(F(Q^{\ast},a),a^{-1}){\chi}^{\mu}_{E}(Q^{\ast})
(\Phi ^{-1}){}^{\beta}_{\mu}(Q^{\ast}){\bar v}^
{\alpha}_{\beta}(a)\frac{\partial}{\partial a^{\alpha}}.
\end{eqnarray*}
Here $F^{C}_{B}(Q,a)\equiv \frac{\partial F^{C}}
{\partial Q^{B}}(Q,a)$, ${\chi}^{\mu}_{E}\equiv
\frac{\partial {\chi}^{\mu}}{\partial Q^{E}}(Q)$,
$(\Phi ^{-1}){}^{\beta}_{\mu}(Q)$ -- the matrix which is inverse to the Faddeev -- Popov matrix:
\[
(\Phi ){}^{\beta}_{\mu}(Q)=K^{A}_{\mu}(Q)
\frac{\partial {\chi}^{\beta}(Q)}{\partial Q^{A}}
\]
($K_{\mu}$ are the Killing vector fields for the Riemannian metric $G_{AB}(Q)$), the matrix  ${\bar v}^{\alpha}_{\beta}(a)$ is the
inverse of
the matrix ${\bar u}^{\alpha}_{\beta}(a)$.  (${\bar u}^{\alpha}_{\beta}(a)$ and  ${ u}^{\alpha}_{\beta}(a)$ are the auxiliary functions 
for the group $\mathcal G$.)

$N^{A}_{C}$ is  the projection operator
($N^{A}_{B}N^{B}_{C}=N^{A}_{C}$) onto the subspace which is   orthogonal to the Killing vector field subspace:
\[
N^{A}_{C}(Q)={\delta}^{A}_{C}-K^{A}_{\alpha }(Q)
(\Phi ^{-1}){}^{\alpha}_{\mu}(Q){\chi}^{\mu}_{C}(Q).
\]
 Being  restricted to the submanifold $\Sigma$, it is equal to $N^A_C(Q^{\ast})$.

After performing the  replacement of the coordinates to a new coordinate basis $(\frac{\partial}{\partial Q^{\ast}{}^A},\frac{\partial}{\partial a^{\alpha}})$,   we come, as in \cite{Storchak_2}, to the following representation for the 
 original metric ${ G}_{ A B}(Q)$ of the manifold $\cal P$:
\begin{equation}
\displaystyle
{\tilde G}_{\cal A\cal B}(Q^{\ast},a)=
\left(
\begin{array}{cc}
G_{CD}(Q^{\ast})(P_{\perp})^{C}_{A}
(P_{\perp})^{D}_{B} & G_{CD}(Q^{\ast})(P_{\perp})^
{D}_{A}K^{C}_{\mu}\bar{u}^{\mu}_{\alpha}(a) \\
G_{CD}(Q^{\ast})(P_{\perp})^
{C}_{A}K^{D}_{\nu}\bar{u}^{\nu}_{\beta}(a) & {\gamma }_{\mu \nu }
(Q^{\ast})\bar{u}_\alpha ^\mu (a)\bar{u}_\beta ^\nu (a)
\end{array}
\right),
\label{1}
\end{equation}
where $G_{CD}(Q^{\ast})\equiv G_{CD}(F(Q^{\ast},e))$, ($e$ is an identity element  of the group $\cal G$),
 ${\gamma}_{\mu \nu}$ is the metric given on the orbit of the group action. It
is defined by the following relation ${\gamma}_{\mu \nu}
=K^{A}_{\mu}G_{AB}K^{B}_{\nu}$.
 $P_{\perp}$ is a projection operator on the tangent 
plane to the submanifold $\Sigma$ given by the gauges $\chi $:
\[
(P_{\perp})^{A}_{B}=\delta ^{A}_{B}-{\chi}^{\alpha}_{B}
(\chi \chi ^{\top})^{-1}{}^{\beta}_{\alpha}(\chi ^
{\top})^{A}_{\beta}.
\]
Here $(\chi ^{\top})^{A}_{\beta}$ is a transposed matrix to the matrix $\chi ^{\nu}_{B}$:
\[
(\chi ^{\top})^{A}_{\mu}=G^{AB}{\gamma}_
{\mu \nu}\chi ^{\nu}_{B}\,\,\, {\gamma}_{\mu \nu}
=K^{A}_{\mu}G_{AB}K^{B}_{\nu}.
\]
The above projection operators have the following properties:
\[
(P_{\perp})^{\tilde A}_{B}N^{C}_{\tilde A}=
(P_{\perp})^{C}_{B},\,\,\,\,\,\,\,\,\,N^{\tilde A}_
{B}(P_{\perp})^{C}_{\tilde A}=N^{C}_{B}.
\]
Note that in the formula (\ref{1}), $K^A_{\mu}$ and $(P_{\bot})^A_B$ are given on $\Sigma$.

The pseudoinverse matrix ${\tilde G}^{\cal A\cal B}(Q^{\ast},a)$ to matrix (\ref{1}) is as follows: 
\begin{equation}
\displaystyle
\left(
\begin{array}{cc}
G^{EF}N^{C}_{E}
N^{D}_{F} & G^{SD}N^C_S{\chi}^{\mu}_D
(\Phi ^{-1})^{\nu}_{\mu}{\bar v}^{\sigma}_{\nu} \\
G^{CB}{\chi}^{\gamma}_C (\Phi ^{-1})^{\beta}_{\gamma}N^D_B
{\bar v}^{\alpha}_{\beta} & G^{CB}
{\chi}^{\gamma}_C (\Phi ^{-1})^{\beta}_{\gamma}
{\chi}^{\mu}_B (\Phi ^{-1})^{\nu}_{\mu}
{\bar v}^{\alpha}_{\beta}{\bar v}^{\sigma}_{\nu}
\end{array}
\right),
\label{2}
\end{equation}
where ${\bar v}^{\sigma}_{\nu}\equiv 
{\bar v}^{\sigma}_{\nu}(a)$ and other components depend on
$Q^{\ast}$.

The pseudoinversion of ${\tilde G}_{\cal B\cal C}$ means that
\begin{eqnarray*}
\displaystyle
{\tilde G}^{\cal A\cal B}{\tilde G}_{\cal B\cal C}=\left(
\begin{array}{cc}
(P_{\perp})^C_B & 0\\
0 & {\delta}^{\alpha}_{\beta}
\end{array}
\right).
\end{eqnarray*}  

\section{The Lagrangian in the horizontal lift basis}
We assume that the considered mechanical system has an invariant Lagrangian which in local coordinates $Q^A$ can be written as follows:
\begin{equation}  
\mathcal L=\frac12 G_{AB}(Q){\dot Q}^A{\dot Q}^B -V(Q),
\label{3}
\end{equation}
where $G_{AB}(Q)$ is an invariant metric (under the action of the group $\mathcal G$) and $V$ is an invariant potential: $V(F(Q,a))=V(Q)$.

The replacement of the coordinates $Q^A$ for $(Q^{\ast}{}^B, a^{\alpha})$, with $Q^A=F^A(Q^{\ast}{}^B, a^{\alpha})$, leads to the transformation of the velocities ${\dot Q}^A(t)$:
\[
{\dot Q}^A(t)\equiv\frac{dQ^A}{dt}=F^A_C\,({P_{\bot}})^C_D\,\frac{d{Q^{\ast}}^D}{dt}+F^A_{\alpha}\,\frac{da^{\alpha}}{dt}.
\]
We note that as an operator, the vector field $\frac{\partial}{\partial
Q^{\ast}{}^{A}}$  is defined in its action on functions by the rule
\[
\left.
\frac{\partial}{\partial Q^{\ast}{}^{A}}
\varphi(Q^{\ast})=(P_{\perp})^{D}_{A}(Q^{\ast})
\frac{\partial \varphi (Q)}{\partial Q^{D}}
\right |_{Q=Q^{\ast}}.
\]
Since $F^A_{\alpha}=F^A_CK^C_{\beta}{\bar u}^{\beta}_{\alpha}$, we can rewrite the right-hand side of the expression for ${\dot Q}^A(t)$ to get
\[
{\dot Q}^A(t) =F^A_C\,  \Bigl(({P_{\bot}})^C_D\,\frac{d{Q^{\ast}}^D}{dt}+K^C_{\beta}(Q^{\ast})\,{\bar u}^{\beta}_{\alpha}(a)\,\frac{da^{\alpha}}{dt}\Bigr)
\]
\[=F^A_C\,  \Bigl(\frac{d{Q^{\ast}}^C}{dt}+K^C_{\beta}(Q^{\ast})\,{\bar u}^{\beta}_{\alpha}(a)\,\frac{da^{\alpha}}{dt}\Bigr).
\]
The last transition has been made by means of the identity $({P_{\bot}})^C_D\,\frac{d{Q^{\ast}}^D}{dt}=\frac{d{Q^{\ast}}^C}{dt}$. The identity is due to the fact that   the velocity vector belongs to the tangent plate to the surface $\Sigma$.

As a result of the replacement of the coordinates we come to the following representation for the Lagrangian:
\begin{equation}
 {\mathcal L}=\frac12G_{CD}(Q^{\ast})\Bigl(\frac{d{Q^{\ast}}^{C}}{dt}+K^C_{\mu}\,{\bar u}^{\mu}_{\alpha}(a)\,\frac{da^{\alpha}}{dt}\Bigr)\Bigl(\frac{d{Q^{\ast}}^{D}}{dt}+K^D_{\nu}\,{\bar u}^{\nu}_{\beta}(a)\,\frac{da^{\beta}}{dt}\Bigr)-V(Q^{\ast}).
\label{4}
\end{equation} 
Before proceeding to the derivation of the Lagrange--Poincar\'{e}  equations, we have to make another replacement of the coordinate  vector fields on the manifold $ {\mathcal P} $. Namely, we change the basis vector fields $(\frac{\partial}{\partial Q^{\ast}{}^A},\frac{\partial}{\partial a^{\alpha}})$ for the horizontal lift basis ($H_A,L_{\alpha}$) introduced in \cite{Storchak_3} as a
  generalisation of  the diagonal lift basis used in \cite{Cho}.

In a new basis, $L_{\alpha}=v^{\mu}_{\alpha}(a)\frac{\partial}{\partial a^{\mu}}$ are the left-invariant vector fields with the
the commutation relations
 \[
[L_{\alpha},L_{\beta}]=c^{\gamma}_{\alpha \beta} L_{\gamma},
\]
where the $c^{\gamma}_{\alpha \beta}$ are the structure constants of the group $\mathcal G$

The horizontal vector fields $H_A$ are determined as follows
\[
 H_A=N^E_A(Q^{\ast}) \left(\frac{\partial}{\partial Q^{\ast}{}^E}-{\tilde {\mathscr A}}^{\alpha}_E\,L_{\alpha}\right),
\]
where ${\tilde{\mathscr A} }^{\alpha}_E(Q^{\ast},a)={\bar{\rho}}^{\alpha}_{\mu}(a)\,{\mathscr A}^{\mu}_E(Q^{\ast})$. 
The matrix ${\bar{\rho}}^{\alpha}_{\mu}$ is inverse to the matrix ${\rho}_{\alpha}^{\beta}$ of the adjoint representation of the group $\cal G$, and  ${\mathscr A}^{\nu}_P={\gamma}^{\nu\mu}K^R_{\mu}\,G_{RP}$ is the mechanical connection  defined in our principal fiber bundle $P(\mathcal M,\mathcal G)$.

The commutation relation of the horizontal vector fields 
\[
 [H_C,H_D]=({\Lambda}^{\gamma}_CN^P_D-{\Lambda}^{\gamma}_DN^P_C)K^{S}_{{\gamma} P}\,H_S-N^E_CN^P_D\,\tilde{\mathcal F}^{\alpha}_{EP}L_{\alpha},
\]
with  ${\Lambda}^{\gamma}_D=({\Phi}^{-1})^{\gamma}_{\mu}\,{\chi}^{\mu}_D$,  
the curvature $\tilde{\mathcal F}^{\alpha}_{EP}$ of the connection ${\tilde{\mathscr A}}$, which is given by
\[
\tilde{\mathcal F}^{\alpha}_{EP}=\displaystyle\frac{\partial}{\partial Q^{\ast}{}^E}\,\tilde{\mathscr A}^{\alpha}_P- 
\frac{\partial}{\partial {Q^{\ast}}^P}\,\tilde{\mathscr A}^{\alpha}_E
+c^{\alpha}_{\nu\sigma}\, \tilde{\mathscr A}^{\nu}_E\,
\tilde{\mathscr A}^{\sigma}_P,
\]
($\tilde{\mathcal F}^{\alpha}_{EP}({Q^{\ast}},a)={\bar{\rho}}^{\alpha}_{\mu}(a)\,{\mathcal F}^{\mu}_{EP}(Q^{\ast})\,$),  
can be rewritten as 
 the commutation relations of the nonholonomic basis:
\begin{equation}
 [H_C,H_D]={\mathscr C}^A_{CD}\,H_A+{\mathscr C}^{\alpha}_{CD}L_{\alpha}
\label{commrelat}
\end{equation}
with the structure constants 
\begin{equation}
{\mathscr C}^A_{CD}=({\Lambda}^{\gamma}_CK^A_{\gamma D}-{\Lambda}^{\gamma}_DK^{A}_{{\gamma} C})
\label{struc_cnst_H}
\end{equation}
and
\begin{equation}
{\mathscr C}^{\alpha}_{CD}=-N^S_CN^P_D\,\tilde{\mathcal F}^{\alpha}_{SP}\,.
\label{struc_cnst_alf}
\end{equation}

In our  basis ($H_A,L_{\alpha}$), $L_{\alpha}$
commutes with $H_A$\,:
\[
[H_A,L_{\alpha}]=0.
\]

And  the  metric  (\ref{1}) has the following diagonal representation: 
\begin{equation}
\displaystyle
{\check G}_{\cal A\cal B}=
\left(
\begin{array}{cc}
G^{\rm H}_{AB} & 0 \\
0 & \tilde{\gamma }_{\alpha \beta }
\end{array}
\right),
\label{metric}
\end{equation}
where
the ``horizontal metric'' $G^{\rm H}$ is defined by 
the projection operator ${\Pi}^{ A}_B={\delta}^A_B-K^A_{\mu}{\gamma}^{\mu \nu}K^D_{\nu}G_{DB}$ as follows:  
$G^{\rm H}_{DC}={\Pi}^{\tilde D}_D\,{\Pi}^{\tilde C}_C\,G_{{\tilde D}{\tilde C}}$.

Note that the projection operator ${\Pi}^{ A}_B$ satisfies the properties: ${\Pi}^{ A}_L N^L_C={\Pi}^{ A}_C$ and ${\Pi}^L_BN^A_L=N^A_B$.

The pseudoinverse matrix ${\check G}^{{\mathcal A}{\mathcal B}}$ to the matrix ${\check G}_{\mathcal B\mathcal C}$  is defined by
the following orthogonality condition:   
\begin{eqnarray*}
\displaystyle
{\check G}^{\mathcal A\mathcal B}{\check G}_{\mathcal B\mathcal C}=\left(
\begin{array}{cc}
N^A_C & 0\\
0 & {\delta}^{\alpha}_{\beta}
\end{array}
\right),
\end{eqnarray*}  
and can be written as
\begin{eqnarray*}
\displaystyle
{\check G}^{\cal A\cal B}=
\left(
\begin{array}{cc}
G^{EF}N^A_EN^B_F & 0 \\
0 & \tilde{\gamma }^{\alpha \beta }
\end{array}
\right).
\end{eqnarray*}

It can be shown that in the horizontal lift  basis ($H_A,L_{\alpha}$),  the Lagrangian (\ref{4}) becomes 
\begin{equation}
 {\hat{\mathcal L}}=\frac12\,G^{\rm H}_{CD}\, {\omega}^C {\omega}^D +\frac12 {\tilde{\gamma}}_{\mu \nu} {\omega}^{\mu} {\omega}^{\nu}-V,
\label{lagrangian}
\end{equation}
where we have introduced the following variables connected with the velocities:
\begin{equation}
{\omega}^E= (P_{\bot})^E_B\, \frac{d Q^{\ast B}}{dt}=\frac{d Q^{\ast E}}{dt}
\label{omega_A}
\end{equation}
and
\begin{equation}
 {\omega}^{\alpha}=u^{\alpha}_{\sigma}\frac{d a^{\sigma}}{dt}+{\omega}^D{\tilde {\mathscr A}}^{\alpha}_{D}.
\label{omega_alf}
\end{equation}
Note also that
\[
 \frac{d a^{\beta}}{dt}=v^{\beta}_{\alpha} \,{\omega}^{\alpha}-{\omega}^Dv^{\beta}_{\alpha} \, {\tilde {\mathscr A}}^{\alpha}_{D}.
\]

\section{The relationship between  partial derivatives of velocities and deformations in the Poincar\'{e} variational principle}
The variational principle proposed Poincar\'{e} for mechanical systems is to use the variations of paths that are associated with independent vector fields provided  they  exist on the configuration space. It is  important that vector fields  may also be  nonholonomic vector fields. 

For example, suppose we have the vector fields $v_1,\ldots,v_n$ on a  some smooth manifold. And these (nonholonomic) vector fields  form a basis. Then, the commutator of the vector fields is expanded over this basis: $[v_i,v_j]=c^k_{ij}(q)v_k$. If we have a some smooth path $q(t)$ on the considered manifold, then the time derivative of a smooth function $f$, given on this path $q(t)$, can be presented as 
\[
 \frac{df(q(t))}{dt}=\frac{\partial f}{\partial q^i} \frac{d q^i}{d t}={\sum}_iv_i(f) {\omega}^i,
\]
where $v_i(f)$ is the directional derivative of $f$ along the vector field $v_i$.
The variables ${\omega}^i$ are called  the quasi-velocities, as they are  linear functions of the velocities $v_i$.

Then, as is done in the usual calculus of variations, it is necessary to introduce the deformation $q(u,t)$ of the path $q(t)$. (These deformations have the standard properties. For the variations with the fixed ends, they are given, for example, in \cite{Arnold}.) 
But now the derivative of the function given on the deformation is calculated  in accordance with  the following formula
\[
 \frac{\partial f(q(u,t))}{\partial u}={\sum}_iv_i(f) {w^i}(u,t),
\]
where the  introduced variations $w^i(u,t)$ are independent within  the time interval $[t_1,t_2]$ which is used for consideration of thevariational problem.  And at the ends of the time interval, they satisfy $w^k(u,t_1)=0$ and $w^k(u,t_2)=0$.

The variation of the functional $F(q(t))$ in this variational calculus is  defined as usual, i.e., as  
\[
 \delta F =\frac{dF(q(u,t))}{du}\Bigl|_{u=0}.
\]

In our case, we are given two sets of the basis vector fields, $\{H_A\}$ and $\{L_{\alpha}\}$, and we know that  vector fields of these sets are independent between themselves: $[H_A,L_{\alpha}]=0$. So, we can apply the Poincar\'{e} variational principle  to the action functional
\begin{equation}
 S=\int_{t_1}^{t_2}{\hat{\mathcal L}}\,dt,
\label{action} 
\end{equation}
where the Lagrangian ${\hat{\mathcal L}}$ is given by (\ref{lagrangian}).

Before applying this variational principle to the functional  (\ref{action}), it is necessary to find  relationship between the derivations of the velocities $\omega ^A$ and $\omega ^{\alpha}$ that are  in the Lagrangian (\ref{lagrangian}) and the variations $w^A$ and $w^{\alpha}$. 

They are follow from the expansion of  the time-derivative of the function $f(Q^{\ast},a)$ in the horizontal lift basis:
\begin{eqnarray}
 \frac{d f(Q^{\ast},a)}{dt}&=&(P_{\bot})^E_B\, \frac{d Q^{\ast B}}{dt} H_E(f)+(P_{\bot})^{D'}_B{\tilde {\mathscr A}}^{\alpha}_{D'}\,L_{\alpha}(f)\frac{d Q^{\ast B}}{dt}+\frac{\partial f}{\partial a^{\alpha}} \frac{d a^{\alpha}}{dt}
\nonumber\\
 &=&{\omega}^E H_E(f)+{\omega}^{\alpha} L_{\alpha}(f),
\label{dif_gen}
\end{eqnarray}
where ${\omega}^E$ and ${\omega}^{\alpha}$ are as in (\ref{omega_A}) and (\ref{omega_alf}), correspondingly, and by $H_E(f)$ we denote the action of the vector field $H_E$ on the function $f$. A similar notation is used for $L_{\alpha}(f)$.

First we consider the relation between the  derivatives  of the functions $\omega ^A$ and the variations  $w^A$. Taking $f=Q^{\ast}{}^A$  in   (\ref{dif_gen}), we get
\[
\frac{d {Q^{\ast}}^A(t)}{dt}=  
{\omega}^E H_E^A(Q^{\ast}(t)),
\]
where
\[
H_E^A(Q^{\ast})\equiv H_E({Q^{\ast}}^A)=  N^A_E(Q^{\ast}).
\]

The previous equality for the time derivative of $Q^{\ast}$ can be generalized to a similar equality for 
the deformation ${Q^{\ast}}^A(u,t)$ of the path ${Q^{\ast}}^A(t)$: 

\begin{equation}
  \frac{d {Q^{\ast}}^A(u,t)}{dt}=  H_E({Q^{\ast}}^A(u,t))\,{\omega}^E(Q^{\ast}(u,t)).
\label{difQ_t}
\end{equation}

On the other hand, for the partial derivative of $Q^{\ast}{}^A(u,t)$ with respect to $u$, we suppose the following equation:
\begin{equation}
  \frac{\partial {Q^{\ast}}^A(u,t)}{\partial u}=
  H_E({Q^{\ast}}^A(u,t))\,{w}^E(Q^{\ast}(u,t)),
\label{difQ_u}
\end{equation}
where we have introduced the variation $w^E(Q^{\ast}(u,t))$.

Now taking the partial derivative of (\ref{difQ_t}) with respect to $u$, we obtain

\begin{eqnarray}
 \frac{\partial}{\partial u}\frac{d {Q^{\ast}}^A(u,t)}{dt}&=&\frac{\partial H^A_E(Q^{\ast})}{\partial {Q^{\ast}}^B}\,\frac{\partial {Q^{\ast}}^B}{\partial u}\, {\omega}^E+H^A_E\,\frac{\partial {\omega}^E(Q^{\ast}(u,t))}{\partial u}
\nonumber\\
&=&\frac{\partial H^A_E}{\partial {Q^{\ast}}^B}\,H^B_Pw^P{\omega}^E+H^A_E\frac{\partial {\omega}^E}{\partial u}.
\label{difQ_t_u}
\end{eqnarray}
But if we peform the differentiation of  (\ref{difQ_u}) with respect to $t$, we get
\begin{equation}
 \frac{\partial}{\partial t}\frac{d {Q^{\ast}}^A(u,t)}{du}=\frac{\partial H^A_E(Q^{\ast})}{\partial {Q^{\ast}}^B}\,
H^B_P{\omega}^Pw^E+H^A_E\frac{\partial {w}^E}{\partial t}.
\label{difQ_u_t}
\end{equation}
Since (\ref{difQ_t_u}) and (\ref{difQ_u_t}) are equal, then  from their equality we obtain the following equation:
\[
\Bigl( \frac{\partial H^A_E}{\partial {Q^{\ast}}^B}H^B_P-\frac{\partial H^A_P}{\partial {Q^{\ast}}^B}H^B_E\Bigr)\,{\omega}^Ew^P+H^A_{E'}\frac{\partial {\omega}^{E'}}{\partial u}-H^A_{E'}\frac{\partial {w}^{E'}}{\partial t}=0.
\]
Taking into account  the commutation relation  between $H_A$ and $H_B$  (\ref{commrelat}), and the expressions for the structure constants (\ref{struc_cnst_H}) and (\ref{struc_cnst_alf}), 
 we come to the following equation:
\begin{equation}
 H^A_R\Bigl(\frac{\partial {\omega}^{R}}{\partial u}-\frac{\partial {w}^{R}}{\partial t}+{\mathcal C}^{R}_{PE}\Bigr)=0.
\label{relat_Q_w}
\end{equation}
(Note  that $H^A_R$ coinsides with the projection operator $N^A_R(Q^{\ast})$.)

The second equation connecting the derivatives of $\omega^{\alpha}$ and the variations $w^\alpha$ can be obtained in the Appendix A. It 
is the equation (\ref{relat_a_w}) and looks as follows:
\[
\frac{\partial {\omega}^{\beta}}{\partial u}=\frac{\partial {w}^{\beta}}{\partial t}+ c^{\beta}_{\alpha ' \mu}\,{\omega}^{\alpha '}w^{\mu}+ N^C_E N^{C'}_P{\tilde {\mathcal F}}^{\beta}_{C' C} \,{\omega}^E w^P.
\]
Now we can proceed to derivation of the Lagrange-Poincar\'{e} equations.

\section{The Lagrange-Poincar\'{e} equation}
To obtain the equations of motions  by means of the  variational principle from an action functional it is necessary to replace the paths in the Lagrangian by their deformations and then to calculate the variation of the action functional. At first we must to calculate the derivative of the functional $S$ with respect to the variable that is connected with the deformation of the paths.
For the functional (\ref{action}) with the Lagrangian (\ref{lagrangian}), this derivative is given  as follows:

\begin{equation}
 \frac{dS}{du}=\int_{t_1}^{t_2}\Bigl(\frac{\partial {\hat{\mathcal L}}}{\partial {\omega}^{C'}}\frac{\partial {\omega}^{C'}}{\partial u}+\frac{\partial {\hat{\mathcal L}}}{\partial {\omega}^{\mu '}}\frac{\partial {\omega}^{\mu '}}{\partial u}+\frac{\partial {\hat{\mathcal L}}}{\partial Q^{\ast}{}^B}\frac{\partial Q^{\ast}{}^B}{\partial u}+\frac{\partial {\hat{\mathcal L}}}{\partial {a}^{\alpha}}\frac{\partial {a}^{\alpha}}{\partial u}\Bigr)dt.
\label{ds_du}
\end{equation}
In (\ref{ds_du}) the first term in the integrand  can be rewritten as
\[
 \frac{\partial {\hat{\mathcal L}}}{\partial {\omega}^{C'}}\frac{\partial {\omega}^{C'}}{\partial u}= 
G^{\rm H}_{CD}\,{\omega}^D\frac{\partial {\omega}^C}{\partial u}.
\]
Our next transformation of this term consists in replacing the partial derivative of $\omega$ with respect to $u$ for the partial derivative of the variation  $w$ with respect to $t$. For this  we make use   of the previously  obtained equation  (\ref{relat_Q_w}) 
\[
 N^{C'}_C\frac{\partial {\omega}^{C}}{\partial u}=N^{C'}_C\frac{\partial {w}^{C}}{\partial t}+N^{C'}_C\,{\mathcal C}^{C}_{EP}\,{\omega}^E w^P,
\]
which we multiply by $G^{\rm H}_{C'D}$. Because of the identity $G^{\rm H}_{CE}N^E_K=G^{\rm H}_{CK}$ we will have
\[
G^{\rm H}_{CD}\frac{\partial {\omega}^{C}}{\partial u}= G^{\rm H}_{CD}\,\Bigl(\frac{\partial {w}^{C}}{\partial t}+\,{\mathcal C}^{C}_{EP}\,{\omega}^E w^P\Bigr).
\]

Therefore, using this expression for the first term  in the integral (\ref{ds_du}) and integrating it by parts, we  obtain
\[
 \Bigl(G^{\rm H}_{CD}\,{\omega}^D\,{w}^{C}\Bigr)\Bigl|^{t_2}_{t_1}
-\int^{t_2}_{t_1}\Bigl(\frac{d}{dt}\Bigl(G^{\rm H}_{CD}\,{\omega}^D\Bigr){w}^{C}-G^{\rm H}_{CD}\,{\omega}^D\,{\mathcal C}^{C}_{EP}\,{\omega}^E w^P\Bigr) dt.
\]
This can also be rewritten as follows:
\[
  \Bigl(G^{\rm H}_{CD}\,{\omega}^D\,{w}^{C}\Bigr)\Bigl|^{t_2}_{t_1}
-\int^{t_2}_{t_1}\Bigl(\frac{d}{dt}\Bigl(\frac{\partial {\hat{\mathcal L}}}{\partial {\omega}^C}\Bigr)-\frac{\partial {\hat{\mathcal L}}}{\partial {\omega}^P}\,{\mathcal C}^{P}_{EC}\,{\omega}^E \Bigr)w^C dt.
\]

Similarly, the second  term can be transformed by the following way
\[
 \frac{\partial {\hat{\mathcal L}}}{\partial {\omega}^{\alpha}}\frac{\partial {\omega}^{\alpha}}{\partial u}={\tilde {\gamma}}_{\alpha\epsilon}{\omega}^{\epsilon}\Bigl(\frac{\partial {w}^{\alpha}}{\partial t}+ c^{\alpha}_{\nu\mu}\,{\omega}^{\nu}w^{\mu}+ N^C_E N^{C'}_P{\tilde {\mathcal F}}^{\alpha}_{C' C} \,{\omega}^E w^P\Bigr).
\]
Here we have used the equation (\ref{relat_a_w}).
Substituting such a representation  in the integral, and then integrating it by parts, we will have
\[
{\tilde {\gamma}}_{\alpha\epsilon}{\omega}^{\epsilon}w^{\alpha}\Bigl|^{t_2}_{t_1}+\int^{t_2}_{t_1}\Bigl(-\frac{d}{dt}\Bigl({\tilde {\gamma}}_{\alpha\epsilon}{\omega}^{\epsilon}\Bigr)w^{\alpha}+{\tilde{\gamma}}_{\alpha\epsilon}{\omega}^{\epsilon}\bigl(c^{\alpha}_{\nu\mu}\,{\omega}^{\nu}w^{\mu}+ N^C_E N^{C'}_P{\tilde {\mathcal F}}^{\alpha}_{C' C} \,{\omega}^E w^P\bigr)\Bigr)dt.
\]
\[
 ={\tilde {\gamma}}_{\alpha\epsilon}{\omega}^{\epsilon}w^{\alpha}\Bigl|^{t_2}_{t_1}+\int^{t_2}_{t_1}\Bigl(-\frac{d}{dt}\Bigl(\frac{\partial {\hat{\mathcal L}}}{\partial {\omega}^{\alpha}}\Bigr)w^{\alpha}+\frac{\partial {\hat{\mathcal L}}}{\partial {\omega}^{\alpha}}\bigl(c^{\alpha}_{\nu\mu}\,{\omega}^{\nu}w^{\mu}+ N^C_E N^{C'}_P{\tilde {\mathcal F}}^{\alpha}_{C' C} \,{\omega}^E w^P\bigr)\Bigr)dt
\]
The last terms in the integrand of (\ref{ds_du}) can be transformed as 
\[
 \frac{\partial {\hat{\mathcal L}}}{\partial Q^{\ast}{}^B}\frac{\partial Q^{\ast}{}^B}{\partial u}+\frac{\partial {\hat{\mathcal L}}}{\partial {a}^{\alpha}}\frac{\partial {a}^{\alpha}}{\partial u}=\frac{\partial {\hat{\mathcal L}}}{\partial Q^{\ast}{}^B}N^B_Ew^E+\frac{\partial {\hat{\mathcal L}}}{\partial {a}^{\alpha}}\Bigl(H_E(a^{\alpha})w^E+L_{\beta}(a^{\alpha})w^{\beta}\Bigr).
\]
Since
\[
 H_E(a^{\alpha})=-N^C_E{\tilde{\mathcal A}}^{\beta}_Cv^{\alpha}_{\beta},\;\;\;\;L_{\beta}(a^{\alpha})=v^{\alpha}_{\beta},
\]
the last terms can also be written as follows:
\[
 H_E({\hat{\mathcal L}})w^E+L_{\alpha}({\hat{\mathcal L}})w^{\alpha}.
\]
Because of the independence of the variations $w^E$ and $w^{\alpha}$ we come to the system of two equations, the Lagrange-Poincar\'{e} equations. The first of the equations, the horizontal equation (after an appropriate changing the notation of the indices) is given by
\begin{equation}
 -\frac{d}{dt}\Bigl(\frac{\partial {\hat{\mathcal L}}}{\partial {\omega}^E}\Bigr)+\frac{\partial {\hat{\mathcal L}}}{\partial {\omega}^P}\,{\mathcal C}^{P}_{CE}\,{\omega}^C+ \frac{\partial {\hat{\mathcal L}}}{\partial {\omega}^{\alpha}}  N^{C'}_E{\tilde {\mathcal F}}^{\alpha}_{C' B} \,{\omega}^B +H_E({\hat{\mathcal L}})=0.
\label{eq_poinc_hor}
\end{equation}
Here we have used the identity $N^B_P{\omega}^P={\omega}^B$.

The second of the Lagrange-Poincar\'{e} equation, the ``vertical'' equation, is as follows:
\begin{equation}
 -\frac{d}{dt}\Bigl(\frac{\partial {\hat{\mathcal L}}}{\partial {\omega}^{\alpha}}\Bigr)+\frac{\partial {\hat{\mathcal L}}}{\partial {\omega}^{\mu}}c^{\mu}_{\nu\alpha}\,{\omega}^{\nu}+L_{\alpha}({\hat{\mathcal L}})=0.
\label{eq_pinc-vert}
\end{equation}
These equations can be rewritten in more explicit form:
\begin{equation}
 -\frac{d}{dt}\Bigl(G^{\rm H}_{E D}{\omega}^D\Bigr)+{\mathcal C}^{R}_{CE}G^{\rm H}_{RD}{\omega}^D{\omega}^C+{\tilde{\gamma}}_{\alpha \beta}\,{\omega}^{\beta} N^{B}_E{\tilde {\mathcal F}}^{\alpha}_{BA} \,{\omega}^A +H_E({\hat{\mathcal L}})=0
\label{eq_poinc_hor_expl}
\end{equation}
and
\begin{equation}
 -\frac{d}{dt}({\tilde{\gamma}}_{\alpha\beta}{\omega}^{\beta})+c^{\mu}_{\nu\alpha}{\tilde{\gamma}}_{\mu\beta}\,{\omega}^{\beta}{\omega}^{\nu}+L_{\alpha}({\hat{\mathcal L}})=0.
\label{eq_poinc_vert_expl}
\end{equation}

It remains to calculate the $H_E({\hat{\mathcal L}})$-terms in the obtained equations. 
We recall that the horizontal vector field $H_E$ is
$H_E=N^S_E(Q^{\ast}) \left(\frac{\partial}{\partial Q^{\ast}{}^S}-{\tilde {\mathscr A}}^{\alpha}_S\,L_{\alpha}\right)$.
Its action on our Lagrangian $\hat{\mathcal L}$ is given by 
\begin{eqnarray}
&&H_E(\hat{\mathcal L})=\frac12N^S_E\frac{\partial}{\partial Q^{\ast}{}^S}\bigl(G^{\rm H}_{AB}\bigr){\omega}^A{\omega}^B+\frac12N^S_E\frac{\partial}{\partial Q^{\ast}{}^S}\bigl({\tilde{\gamma}}_{\nu\epsilon}\bigr){\omega}^{\nu}
{\omega}^{\epsilon}
\nonumber\\
&& \;\;\;\;-\frac12N^S_E{\tilde {\mathscr A}}^{\alpha}_S\,L_{\alpha}({\tilde{\gamma}}_{\nu\epsilon}) {\omega}^{\nu} {\omega}^{\epsilon}-N^S_E\frac{\partial}{\partial Q^{\ast}{}^S}\bigl(V(Q^{\ast})\bigr).
\label{H_E(L)}
\end{eqnarray}
First note that the second and the third terms of $H_E(\hat{\mathcal L})$ can be rewritten as
\[
 \frac12N^S_E({\tilde{\mathscr D}}_S{\tilde{\gamma}}_{\nu\epsilon}){\omega}^{\nu} {\omega}^{\epsilon}=-\frac12N^S_E({\mathscr D}_S{{\gamma}}^{\kappa\sigma}){\gamma}_{\alpha\sigma}{\rho}^{\alpha}_{\epsilon}{\gamma}_{\kappa\beta}{\rho}^{\beta}_{\nu}\,{\omega}^{\nu} {\omega}^{\epsilon}.
\]
Then we rewrite the equation (\ref{eq_poinc_hor_expl}) in the following form:
\begin{equation}
 -G^{\rm H}_{E D}\frac{d}{dt}\,{\omega}^D-\frac{d}{dt}\Bigl(G^{\rm H}_{E D}\Bigr){\omega}^D+
{\mathcal C}^{R}_{CE}G^{\rm H}_{RD}{\omega}^D{\omega}^C+{\tilde{\gamma}}_{\alpha \beta}\,{\omega}^{\beta} N^{B}_E{\tilde {\mathcal F}}^{\alpha}_{BA} \,{\omega}^A +H_E({\hat{\mathcal L}})=0.
\label{eq_poinc_hor_expl_2}
\end{equation}
Note that the derivation of $G^{\rm H}_{E D}$ with respect to time can be performed in accordance with the following rule:
\[
\frac{d}{dt}\Bigl(G^{\rm H}_{E D}(Q^{\ast})\Bigr)=(P_{\bot})^M_S\Bigl(\frac{\partial G^{\rm H}_{E D}(Q)}{\partial Q^M}\Bigr)\Bigr|_{Q=Q^{\ast}}{\omega}^S=\frac{\partial G^{\rm H}_{E D}}{\partial Q^{\ast}{}^M}\,{\omega}^M.
\]
(Also note that here we have used  the identity $(P_{\bot})^M_S{\omega}^S={\omega}^M$.)

The first term of $H_E({\hat{\mathcal L}})$ can be presented as

\[
 N^S_E\frac{\partial}{\partial Q^{\ast}{}^S}\Bigl(\frac12 G^{\rm H}_{A B}{\omega}^A{\omega}^B\Bigr)=\frac12 N^S_E(P_{\bot})^M_S\frac{\partial G^{\rm H}_{A B}(Q)}{\partial Q^M}\Bigr|_{Q=Q^{\ast}}{\omega}^A{\omega}^B
\]
\[\;\;\;\;\;\;\;\;\;\;\;\;=\frac12 N^M_E\frac{\partial G^{\rm H}_{A B}}{\partial Q^{\ast}{}^M}\,{\omega}^A{\omega}^B.
\]
We combine it with the term of horizontal equation (\ref{eq_poinc_hor_expl_2})   which contains  the derivative with respect to time of $G^{\rm H}_{E D}$:
\[
(-G^{\rm H}_{E D,M} {\omega}^M{\omega}^D+\frac12 N^M_EG^{\rm H}_{A B,M}{\omega}^A{\omega}^B),
\]
or
\[
 (-G^{\rm H}_{E A,B} +\frac12 N^M_EG^{\rm H}_{A B,M}){\omega}^A{\omega}^B.
\]
But
\[
 N^M_EG^{\rm H}_{MA ,B}=G^{\rm H}_{EA ,B}-K^M_{\alpha}{\Lambda}^{\alpha}_E\,G^{\rm H}_{MA ,B}.
\]
So
\[
 G^{\rm H}_{EA ,B}=N^M_EG^{\rm H}_{MA ,B}+K^M_{\alpha}{\Lambda}^{\alpha}_EG^{\rm H}_{MA ,B}.
\]
Using this representation for $G^{\rm H}_{EA ,B}$, we come to 
\[
-\Bigl(N^M_E(\frac12)(G^{\rm H}_{MA ,B}+G^{\rm H}_{MB ,A}-G^{\rm H}_{AB ,M})+K^M_{\alpha}{\Lambda}^{\alpha}_EG^{\rm H}_{MA ,B}\Bigr){\omega}^A{\omega}^B,
\]
or
\[
 -\Bigl(N^M_EG^{\rm H}_{M P}{}^{\rm H}{\Gamma}^P_{AB}+K^M_{\alpha}{\Lambda}^{\alpha}_EG^{\rm H}_{MA ,B}\Bigr){\omega}^A{\omega}^B.
\]
Notice that the second term of this expression is mutually concealed with the ``${\mathcal C}^{R}_{CE}$ - term'' of 
(\ref{eq_poinc_hor_expl_2}). It can be done as follows.

The two terms can be rewritten as
\[
 (G^{\rm H}_{RA}{\mathcal C}^{R}_{BE}-K^M_{\alpha}{\Lambda}^{\alpha}_E\,G^{\rm H}_{MA ,B}){\omega}^A{\omega}^B.
\]

 Because of ${\Lambda}^{\gamma}_B{\omega}^B=0$, 
\[
 {\mathcal C}^{R}_{BE}{\omega}^A{\omega}^B=({\Lambda}^{\gamma}_BK^R_{\gamma E}-{\Lambda}^{\gamma}_EK^R_{\gamma B}){\omega}^A{\omega}^B=-{\Lambda}^{\gamma}_EK^R_{\gamma B}{\omega}^A{\omega}^B.
\]
And we come to 
\[
 {\Lambda}^{\gamma}_E(-G^{\rm H}_{RA}K^R_{\gamma B}-K^R_{\gamma }G^{\rm H}_{RA,B}){\omega}^A{\omega}^B.
\]
If one takes the partial derivative of the following equality
\[
 K^R_{\gamma }G^{\rm H}_{RA}=0,
\]
one  obtains
\[
 \frac{\partial}{\partial Q^{\ast}{}^B}(K^R_{\gamma }G^{\rm H}_{RA})=(P_{\bot})^S_B(K^R_{\gamma S}G^{\rm H}_{RA}+K^R_{\gamma }G^{\rm H}_{RA,S})=0.
\]
Since $(P_{\bot})^S_B{\omega}^B={\omega}^S$, it follows that
\[
 K^R_{\gamma }G^{\rm H}_{RA,S}{\omega}^S=-K^R_{\gamma S}G^{\rm H}_{RA}{\omega}^S,
\]
and  we  have
\[
 {\Lambda}^{\gamma}_E(-G^{\rm H}_{RA}K^R_{\gamma B}+K^R_{\gamma B}G^{\rm H}_{RA}){\omega}^A{\omega}^B=0.
\]

As a result of our transformation we get
\begin{eqnarray*}
&&-G^{\rm H}_{E D}\frac{d}{dt}\,{\omega}^D-N^M_EG^{\rm H}_{M P}{}^{\rm H}{\Gamma}^P_{AB}\,{\omega}^A{\omega}^B+{\tilde{\gamma}}_{\alpha \beta}\,{\omega}^{\beta} N^{B}_E{\tilde {\mathcal F}}^{\alpha}_{BA} \,{\omega}^A\nonumber\\
&&-\frac12N^S_E({\mathscr D}_S{{\gamma}}^{\kappa\sigma}){\gamma}_{\alpha\sigma}{\rho}^{\alpha}_{\epsilon}{\gamma}_{\kappa\beta}{\rho}^{\beta}_{\nu}\,{\omega}^{\nu} {\omega}^{\epsilon}-N^S_E\frac{\partial}{\partial Q^{\ast}{}^S}V(Q^{\ast})=0.
\end{eqnarray*}

Multiplying this equation by $G^{L'E}N^L_{L'}$ and using the identities $G^{L'E}G^{\rm H}_{E D}={\Pi}^{L'}_D$ and ${\Pi}^{L'}_DN^L_{L'}=N^L_D$, we obtain
\begin{eqnarray}
&&-N^L_D\Bigl(\frac{d}{dt}\,{\omega}^D+{}^{\rm H}{\Gamma}^D_{AB}\,{\omega}^A{\omega}^B+G^{DE}N^B_E{\tilde{\gamma}}_{\alpha \beta}\,{\omega}^{\beta} {\tilde {\mathcal F}}^{\alpha}_{AB} \,{\omega}^A\Bigr.\nonumber\\
&&+\Bigl.\frac12G^{DE}N^S_E({\mathscr D}_S{{\gamma}}^{\kappa\sigma}){\gamma}_{\alpha\sigma}{\rho}^{\alpha}_{\epsilon}{\gamma}_{\kappa\beta}{\rho}^{\beta}_{\nu}\,{\omega}^{\nu} {\omega}^{\epsilon}+G^{DE}N^S_E\frac{\partial}{\partial Q^{\ast}{}^S}V\Bigr)=0.
\end{eqnarray}
In this equation it  is convenient to introduce a new variable $p_\sigma$ instead of ${\omega}^{\nu}$:
\[
 p_\sigma={\gamma}_{\alpha\sigma}{\rho}^{\alpha}_{\epsilon}{\omega}^{\epsilon}.
\]
Then the obtained equation is rewritten as
 \begin{eqnarray}
&&-N^L_D\Bigl(\frac{d}{dt}\,{\omega}^D+{}^{\rm H}{\Gamma}^D_{AB}\,{\omega}^A{\omega}^B+G^{DE}N^B_E { {\mathcal F}}^{\mu}_{AB}\,p_{\mu} \,{\omega}^A{\omega}^B+\Bigr.\nonumber\\
&&\;\;\;\;\;\;\;\;\;\;+\Bigl.\frac12G^{DE}N^S_E({\mathscr D}_S{{\gamma}}^{\kappa\sigma})p_{\kappa}p_{\sigma}+G^{DE}N^S_E\frac{\partial}{\partial Q^{\ast}{}^S}V(Q^{\ast})\Bigr)=0.
\label{final_hor_eq}
\end{eqnarray}

The second Lagrange-Poincar\'{e} equation is
\[
 -\frac{d}{dt}({\tilde{\gamma}}_{\alpha\epsilon}{\omega}^{\epsilon})+c^{\mu}_{\nu\alpha}\,{\tilde{\gamma}}_{\mu\epsilon}{\omega}^{\epsilon}{\omega}^{\nu}+L_{\alpha}({\hat{\mathcal L}})=0,
\]
in which
\[
 L_{\alpha}({\hat{\mathcal L}})=\frac12(c^{\mu}_{\alpha\nu}{\tilde\gamma}_{\mu\epsilon}+c^{\mu}_{\alpha\epsilon}
{\tilde\gamma}_{\nu\mu})\,{\omega}^{\nu}{\omega}^{\epsilon}=c^{\mu}_{\alpha\nu}{\tilde\gamma}_{\mu\epsilon}\,{\omega}^{\nu}{\omega}^{\epsilon}.
\]
It follows that the second term of the equation  cancels   the third one.

The first term can be rewritten as follows:
\[
 \frac{d}{dt}({\rho}^{\nu}_{\alpha}p_{\nu})={\rho}^{\nu}_{\alpha}\frac{d}{dt}p_{\nu}+\frac{\partial {\rho}^{\nu}_{\alpha}}{\partial a^{\mu}}\Bigl(\frac{d a^{\mu}}{d t}\Bigr)p_{\nu}.
\]
But $$\frac{d a^{\mu}}{d t}=v^{\mu}_{\sigma}{\omega}^{\sigma}-{\omega}^C{\tilde{\mathscr A}}^{\kappa}_Cv^{\mu}_{\kappa}.$$
Substituting this expression for $\frac{d a^{\mu}}{d t}$, we 
 get 
\[
 {\rho}^{\nu}_{\alpha}\frac{d}{dt}p_{\nu}+v^{\mu}_{\sigma}\frac{\partial {\rho}^{\nu}_{\alpha}}{\partial a^{\mu}}\,{\omega}^{\sigma}p_\nu-v^{\mu}_{\kappa}\frac{\partial {\rho}^{\nu}_{\alpha}}{\partial a^{\mu}}{\tilde{\mathscr A}}^{\kappa}_C{\omega}^Cp_\nu =0.
\]
In this equation, we have $L_{\sigma}({\rho}^{\nu}_{\alpha})=c^{\varphi}_{\sigma \alpha}{\rho}^{\nu}_{\varphi}$ and $c^{\varphi}_{\sigma \alpha}{\rho}^{\nu}_{\varphi}=c^\nu_{\epsilon \delta}{\rho}^\delta_{\alpha}{\rho}^\epsilon_{\sigma}$. 
So we can multiply the equation by $\bar{\rho}^\alpha_\beta$ to obtain
\[
 \frac{d}{dt}p_{\beta}+c^\nu_{\epsilon \beta}{\rho}^\epsilon_{\sigma}{\omega}^{\sigma}p_\nu-c^\nu_{\epsilon \beta}{\rho}^\epsilon_{\kappa}{\tilde{\mathscr A}}^{\kappa}_C{\omega}^Cp_\nu=0.
\]
The equation may be rewritten in the final following form:
\begin{equation}
 \frac{d}{dt}p_{\beta}+c^\nu_{\epsilon \beta}{\gamma}^{\epsilon \varphi}p_{\varphi}p_\nu-c^\nu_{\mu \beta}{{\mathscr A}}^{\mu}_C{\omega}^Cp_\nu=0.
\label{final_vert_eq}
\end{equation}
Thus, the Lagrange-Poincar\'{e} equations for our Lagrangian (\ref{lagrangian}) are given by the horizontal equation (\ref{final_hor_eq}) and the vertical equation (\ref{final_vert_eq}).

As an important consequence of the obtained equations    are  the equations for 
 the  relative equilibrium: 
\begin{eqnarray}
&&G^{DE}N^L_DN^S_E\Bigl(\frac12({\mathscr D}_S{{\gamma}}^{\kappa\sigma})p_{\kappa}p_{\sigma}+\frac{\partial}{\partial Q^{\ast}{}^S}V(Q^{\ast})\Bigr)=0,\nonumber\\ 
&&c^\nu_{\epsilon \beta}{\gamma}^{\epsilon \varphi}p_{\varphi}p_\nu=0.
\end{eqnarray}

\section{Conclusion}
We have obtained the local Lagrange-Poincar\^{e} equations defined in some chart of the principle fibre bundle. If the principle bundle is a trivial one, these equations may be considered as a global equations. In general, to determine the equations on the whole manifold  
it is necessary to know how they are changed under  the  transition from one chart to another. 

Earlier, these equations were derived in our paper \cite{Storchak_wong_eq}  as the geodesic equations in the horizontal lift basis on a manifold. The horizontal equations of this paper coinsides with the one from our previous work, but the vertical equations are slightly
differed. This is connected with the different definition of the variable $p$. The previous definition was without the mechanical connection. 

Note that the horizontal Lagrange-Poincar\^{e} equation is  known by the name Wong's equation \cite{Wong} in physical literature. 
But in reduction theory, the mechanical connection in the Lagrange-Poincar\^{e} equations  is not an arbitrary as in the Wong's equations, but is determined by the geometry of the reduced problem. Therefore, it would be interested in to study the  questions related to the behaviour of the  Lagrange-Poincar\^{e} equation  in fields theories. 

Note also that generalization  of the equation of the relative equilibrium to the field-theoretical equations could be  useful for studies of the possible stable configurations that can be used in perturbative calculations.

\appendix
\section*{Appendix A}
\section*{Relationship between derivatives of  velocities $\omega ^{\alpha}$ and   variations $w^{\alpha}$ }
\setcounter{equation}{0}
\def\theequation{A.\arabic{equation}}

The velocity $d a^\alpha/{dt}$ can be expressed in terms of the vector fields of the horizontal lift basis:
\[
\frac{d a^{\alpha}}{dt}={\omega}^E H_E(a^{\alpha})+{\omega}^{\beta '}L_{\beta '}(a^{\alpha}),
\]
where
\[
 H_E(a^{\alpha})=-N^C_E{\tilde {\mathscr A}}^{\beta'}_{C}v^{\alpha}_{\beta'},\;\;\;\;\;
 {\tilde {\mathscr A}}^{\beta}_{C}(Q^{\ast},a)={\bar {\rho}}^{\beta}_{\mu}(a){{\mathscr A}}^{\mu}_{C}(Q^{\ast}),
\]
 and  $
 L_{\mu}(a^{\beta})=v^{\beta}_{\mu}.
$

(Note that we have the following action of the vector field $L_{\alpha}$ on the matrix of the adjoint representation of the group Lie:
$
 L_{\mu}\,{\bar {\rho}}^{\nu}_{\epsilon}(a)=-c^{\nu}_{\mu \kappa}\,{\bar {\rho}}^{\kappa}_{\epsilon}(a).
$)

The velocities of deformations $a^{\alpha}(u,t)$ of the path $a^{\alpha}(t)$ have  a similar expansion over the basis ($H_A, L_\alpha)$, which is  taken now on  deformed paths:
\begin{equation}
  \frac{d a^{\alpha}(u,t)}{dt}= H_E(a^{\alpha}(u,t)){\omega}^E(Q^{\ast}(u,t))+L_{\beta '}(a^{\alpha}(u,t))\,{\omega}^{\beta '}(Q^{\ast}(u,t),a(u,t)).
\label{difa_t}
\end{equation}

Assuming the same structure for the partial derivative of $a^{\alpha}(u,t)$ with respect to $u$ and introducing the variations, instead of the velocities ${\omega}^A$ and ${\omega}^{\beta}$, we write this partial derivative of $a^\alpha$ as
\begin{equation}
  \frac{\partial a^{\alpha}(u,t)}{\partial u}= H_E(a^{\alpha}(u,t)){w}^E(Q^{\ast}(u,t))+L_{\beta '}(a^{\alpha}(u,t))\,{w}^{\beta '}(Q^{\ast}(u,t),a(u,t)).
\label{difa_u}
\end{equation}

Taking the partial derivative of (\ref{difa_t}) with respect of $u$, we get
\begin{eqnarray}
&&\!\!\!\!\!\!\!\frac{\partial}{\partial u}\frac{d a^{\alpha}}{dt}=
\frac{\partial H_E(a^{\alpha})}{\partial {Q^{\ast}}^B}H^B_P(Q^{\ast})w^P{\omega}^E
+\frac{\partial H_E(a^{\alpha})}{\partial a^{\beta}}
\Bigl(H_P(a^{\beta})w^P
+L_{\alpha '}(a^{\beta})
w^{\alpha '}\Bigr){\omega}^E
\nonumber\\
&&\!\!\!\!\!\!\!\!\!\!\!+H_E(a^{\alpha})\frac{\partial {\omega}^E}{\partial u}+\frac{\partial L_{\alpha '}(a^{\alpha})}{\partial a^{\beta}}\Bigl(H_P(a^{\beta})w^P
+L_{\mu}(a^{\beta})w^{\mu}\Bigr){\omega}^{\alpha '}+L_{\mu}(a^{\alpha})\frac{\partial {\omega}^{\mu}}{\partial u}.
\label{difa_t_u}
\end{eqnarray}
The partial derivative of (\ref{difa_u}) with respect to $t$ gives us
\begin{eqnarray}
&&\!\!\!\!\!\!\!\frac{d}{dt}\frac{\partial a^{\alpha}}{\partial u}=
\frac{\partial H_P(a^{\alpha})}{\partial {Q^{\ast}}^B}H^B_E(Q^{\ast}){\omega}^Ew^P
+\frac{\partial H_P(a^{\alpha})}{\partial a^{\beta}}
\Bigl(H_E(a^{\beta}){\omega}^E
+L_{\mu}(a^{\beta})
{\omega}^{\mu}\Bigr){w}^P
\nonumber\\
&&\!\!\!\!\!\!\!\!\!\!\!+H_E(a^{\alpha})\frac{\partial {w}^E}{\partial t}+\frac{\partial L_{\mu}(a^{\alpha})}{\partial a^{\beta}}\Bigl(H_P(a^{\beta}){\omega}^P
+L_{\alpha '}(a^{\beta}){\omega} ^{\alpha '}\Bigr){w}^{\mu}+L_{\mu}(a^{\alpha})\frac{\partial {w}^{\mu}}{\partial t}.
\label{difa_u_t}
\end{eqnarray}

The equality of (\ref{difa_t_u}) and (\ref{difa_u_t}) leads us to the following equation:
\[
\Bigl(H^B_P(Q^{\ast}) \frac{\partial H_E(a^{\alpha})}{\partial {Q^{\ast}}^B}-H^B_E(Q^{\ast})\frac{\partial H_P(a^{\alpha})}{\partial {Q^{\ast}}^B}\Bigr){\omega}^Ew^P
\]
\[
+\Bigl(H_P(a^{\beta})\frac{\partial H_E(a^{\alpha})}{\partial a^{\beta}}-H_E(a^{\beta})\frac{\partial H_P(a^{\alpha})}{\partial a^{\beta}}\Bigr){\omega}^Ew^P
\]
\[
+\frac{\partial H_E(a^{\alpha})}{\partial a^{\beta}}L_{\mu}(a^{\beta})(w^{\mu}{\omega}^E- {\omega}^{\mu}w^E)+\Bigl(H_E(a^{\alpha})\frac{\partial {\omega}^{E}}{\partial u}-H_E(a^{\alpha})\frac{\partial {w}^{E}}{\partial t}\Bigr)
\]
\[
 +\Bigl(H_P(a^{\beta})\frac{\partial L_{\alpha '}(a^{\alpha})}{\partial a^{\beta}}w^P{\omega}^{\alpha '}-H_P(a^{\beta})\frac{\partial L_{\mu}(a^{\alpha})}{\partial a^{\beta}}{\omega}^Pw^{\mu}\Bigr)
\]
\[
 +\frac{\partial L_{\alpha '}(a^{\alpha})}{\partial a^{\beta}}L_{\mu}(a^{\beta})w^{\mu}{\omega}^{\alpha '}-\frac{\partial L_{\mu}(a^{\alpha})}{\partial a^{\beta}}L_{\alpha '}(a^{\beta}){\omega}^{\alpha '}w^{\mu}
\]
\[
 +L_{\mu}(a^{\alpha})\frac{\partial {\omega}^{\mu}}{\partial u}-L_{\mu}(a^{\alpha})\frac{\partial {w}^{\mu}}{\partial t}.
\]
After fulfilling the corresponding changes, these nine terms of the obtained equation can be rewritten to give

1+2 
\[
 N^C_E N^{C'}_P{\tilde {\mathcal F}}^{\beta}_{C C'}\,v^{\alpha}_{\beta} \,{\omega}^E w^P+\Bigl(N^C_E\frac{\partial}{\partial Q^{\ast}{}^C}\Bigl(N^{C'}_P\Bigr){\tilde {\mathscr A}}^{\beta}_{C'}-N^{C'}_P\frac{\partial}{\partial Q^{\ast}{}^{C'}}\Bigl(N^{C}_E\Bigr){\tilde {\mathscr A}}^{\beta}_{C}\Bigr){\omega}^E w^P
\]

3

\[
 -N^C_E{{\mathscr A}}^{\nu}_{C}v^{\beta}_{\mu}\frac{\partial}{\partial a^{\beta}}\bigl({\bar v}^{\alpha}_{\nu}\bigr)w^{\mu}{\omega}^E+
N^{C'}_E{{\mathscr A}}^{\nu}_{C'}v^{\beta}_{\mu}\frac{\partial}{\partial a^{\beta}}\bigl({\bar v}^{\alpha}_{\nu}\bigr){\omega}^{\mu}{w}^E
\]

4

\[
 -N^C_E{\tilde {\mathscr A}}^{\beta}_{C}v^{\alpha}_{\beta}\Bigl(\frac{\partial {\omega}^E}{\partial u}-\frac{\partial {w}^E}{\partial t}\Bigr)
\]

5

\[
 -N^C_P{{\mathscr A}}^{\nu}_{C}{\bar v}^{\beta}_{\nu}\frac{\partial}{\partial a^{\beta}}\bigl({ v}^{\alpha}_{\alpha '}\bigr)w^{P}
{\omega}^{\alpha '}+
N^{C}_P{{\mathscr A}}^{\nu}_{C}{\bar v}^{\beta}_{\nu}\frac{\partial}{\partial a^{\beta}}\bigl({\bar v}^{\alpha}_{\mu}\bigr){\omega}^{P}{w}^{\mu} 
\]

6+7
\[
 c^{\sigma '}_{\mu \alpha '}v^{\alpha}_{\sigma '}\,w^{\mu}\,{\omega}^{\alpha '}
\]

8+9

\[
 v^{\alpha}_{\mu}\frac{\partial {\omega}^{\mu}}{\partial u}-v^{\alpha}_{\mu}\frac{\partial {w}^{\mu}}{\partial t}.
\]
Note that it can be shown that the sum of the third  and the fifth  terms  is equal to zero. 

By making use of the equation (\ref{relat_Q_w}) of the main text of the paper 
in the fourth term,  one can show that the resulting  expression cancels with the second term standing in the sum of the first and second terms, i.e., in (1+2)-term. 

Finally, after all transformations one can obtains the following equation which relates the partial derivative of ${\omega}^{\beta}$ and the partial derivative of the variation ${w}^{\beta}$:
\begin{equation}
\frac{\partial {\omega}^{\beta}}{\partial u}=\frac{\partial {w}^{\beta}}{\partial t}+ c^{\beta}_{\alpha ' \mu}\,{\omega}^{\alpha '}w^{\mu}+ N^C_E N^{C'}_P{\tilde {\mathcal F}}^{\beta}_{C' C} \,{\omega}^E w^P.
\label{relat_a_w}
\end{equation}

\end{document}